# *The Supernova of 1054AD, the Armenian chronicle of Hetum, and Cronaca Rampona*

V.G.Gurzadyan

Center for Cosmology and Astrophysics, Alikhanian National Laboratory,

Yerevan, Armenia

The rareness of nearby supernovae ensures particular value to the historic records for determination of their light curves. We provide the translation of 13th century Armenian chronicle of Hetum, which by its unexpected association to *Cronaca Rampona* and other chronicles can influence the debates whether there are reliable European records of the supernova of 1054 AD, as well as the analysis of the records vs the conjunction with the Moon and their role in assigning of the Type I or II to that supernova.

Historical records of astronomical phenomena are often useful for dating historical events[1], and occasionally also for understanding those phenomena themselves. This letter concerns the famous supernova of 1054AD, the progenitor of the Crab Nebula (Messier 1) in Taurus, described in Chinese, Japanese and Arabic records. Although those sources mention that the event was seen even during the daytime, along some other details, there is no accurate information on the date of its first appearance, and hence any new record is of a particular value. Whether there are reliable recordings in Europe is still a matter of debate; the presence of such sources was argued by Collins *et al* [2], while the contrary is concluded by Stephenson & Green[3]. Both groups of authors discuss the *Cronaca Rampona* (first published by Sorbelli[4] ) and other chronicles, along with the Armenian chronicle of Hetum, a 13th-Century author, known as *Hetum Patmich* (Historian). Hetum's chronicle was first published by Hakobian[5] and has been discussed since 1960s by Tumanian and Astapovich[6] initially in the context of a meteor events, and by Collins *et al*[2] in relation to 1054 supernova and other chronicles. Note that Armenian medieval sources do contain records of numerous astronomical phenomena (see, *e.g.* ref.7 ). Stephenson & Green[3] note that "the renewed examination of the entry in the Armenian chronicles is desirable."

I have translated that entry from the manuscript No.1898 (see the image in Fig. 1) in *Matenadaran*, the Institute of Ancient Manuscripts, in Yerevan:

*1048AD. It was the 5th year, 2nd month, 6th day of Pope Leo in Rome. Robert Kijart arrived in Rome and sieged the Tiburtina town. There was starvation over the whole world. That year a bright star appeared within the circle of the Moon, the Moon was new, on May 14$^{th}$, in the first part of the night.*

Here, Pope St.Leo IX (1049- 19 April, 1054), Tiburtina, then a suburb of Rome, are mentioned. The 'circle of the Moon' refers to the Moon's disk, hence, describing a conjunction with an impression of appearance of the 'bright star' within the disk (for discussion see refs, 2 and 3).

An unexpected conclusion can be drawn from the translated passage: the similarity of its wording, including the mentioning of Tiburtina, with that of *Cronaca Rampona*[3], leaves no doubt that we deal with an *identical* source for both of these chronicles. Hetum's chronicle spanning the history from 1AD to 1294AD, as the author himself mentions, includes compilations from French[5]; its source can be not a single one, among those e.g. the 13th-Century chronicle of Martinus Polonus (see [8]).

Thus, we conclude that:

(a) Hetum's chronicle and *Cronaca Rampona* must have a common source and a search for it is of a particular importance;
(b) The conjunctions of the Moon have to be analyzed not for an observer in Armenia (as attempted in refs.2 and 3) but in Western Europe.

The clarification of these issues and of a number of related ones (e.g. the role of near-horizon refraction in the appearance of the conjunction, i.e. linked to the geographic location) and the joint analysis of various copies/translations will reveal whether the sought original source we seek really indicates a European observation of supernova of 1054AD.

Fig. 1. The entry of Hetum’s chronicle with the astronomical record; from the Armenian manuscript No.1898 in Matenadaran, Yerevan (photo by V.G.Gurzadyan).

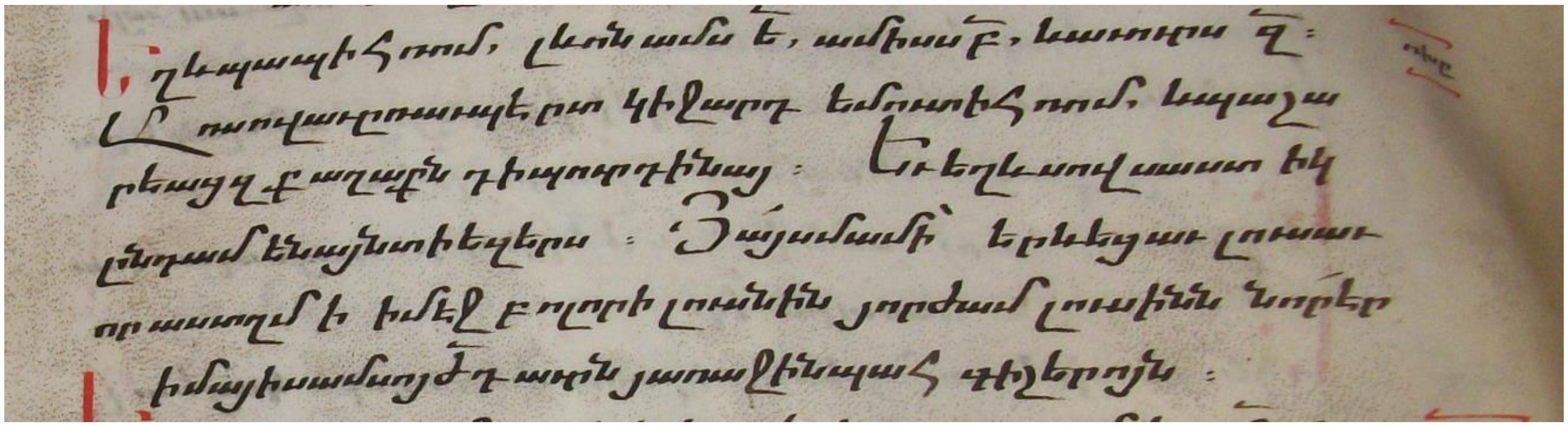